\documentclass{bmvc2k}

\usepackage{amssymb}
\usepackage{booktabs}
\usepackage{graphbox}

\usepackage{multirow}
\usepackage{tabularx}

\usepackage{algorithm}
\usepackage{setspace}
\usepackage[noend]{algpseudocode}

\usepackage{color}

\hypersetup{
pdfborder = {0 0 0}
}

\title{Recurrence-in-Recurrence Networks for Video Deblurring}

\addauthor{Joonkyu Park}{jkpark0825@snu.ac.kr}{1}
\addauthor{Seungjun Nah}{seungjun.nah@gmail.com}{1}
\addauthor{Kyoung Mu Lee}{kyoungmu@snu.ac.kr}{1}

\addinstitution{
 Department of ECE, ASRI\\
 Seoul National University\\
 Korea
}

\runninghead{Park, Nah, Lee}{Recurrence-in-Recurrence Networks}

\def\etal{\emph{et al}\bmvaOneDot}

\newcommand {\revision}[1]{{\color{black}#1}\normalfont}

\newcommand{\norm}[1]{\|{#1}\|}

\newcommand{\figcspace}{\vspace{3mm}}
\newcommand{\figspace}{\vspace{-5mm}}
\newcommand{\tabcspace}{\vspace{3mm}}
\newcommand{\tabspace}{\vspace{-5mm}}

\begin{document}

\maketitle

\begin{abstract}

State-of-the-art video deblurring methods often adopt recurrent neural networks to model the temporal dependency between the frames.
While the hidden states play key role in delivering information to the next frame, abrupt motion blur tend to weaken the relevance in the neighbor frames.
In this paper, we propose recurrence-in-recurrence network architecture to cope with the limitations of short-ranged memory.
We employ additional recurrent units inside the RNN cell.
First, we employ inner-recurrence module~(IRM) to manage the long-ranged dependency in a sequence.
IRM learns to keep track of the cell memory and provides complementary information to find the deblurred frames.
Second, we adopt an attention-based temporal blending strategy to extract the necessary part of the information in the local neighborhood.
The adpative temporal blending~(ATB) can either attenuate or amplify the features by the spatial attention.
Our extensive experimental results and analysis validate the effectiveness of IRM and ATB on various RNN architectures.

\end{abstract}

\section{Introduction}
\label{sec:intro}

Videos often suffer from motion blur caused by the relative motions between the camera and the objects.
From the motion blur that varies both spatially and temporally, recovering the sharp and rich details is a challenging task.
Such complex blur makes the scene hard to be identified, hindering the following scene recognition algorithms to be applied in practice.
As motion blur is one of the  most common artifacts in videos, many efforts were made to address the video deblurring problem.

From the temporal variation of scenes, conventional video deblurring methods tried to extract the motion information to remove it from the scenes.
Motion flow was used to infer the blur trajectories in the frames~\cite{Kim_2015_CVPR,kim2017dynamic} in the joint optimization process with the latent frames.
In \cite{Su_2017_CVPR,Kim_2018_ECCV,Pan_2020_CVPR}, the displacement of the neighboring frames are modeled by optical flow to help the learning of the following neural network models.
The frames are aligned by warping from the estimated flow so that the relevant information could be better aggregated in the learning process.
However, trying to find such point-to-point correspondences from blurry frames are prone to be erroneous and cause misalignment errors.

Recurrent neural networks, on the other hand, try to handle the temporal propagation of scenes in an implicit manner.
Rather than explicitly finding the motion flow, the hidden states convey the information from the past frames to the future frames.
Thus, the way to handle hidden states has played key roles in designing recurrent networks for video deblurring~\cite{Kim_2017_ICCV,Wieschollek_2017_ICCV,Nah_2019_CVPR,Zhou_2019_ICCV,zhong2020efficient}.

As the hidden states are obtained from the past frame but blindly to the current frame, the hidden state should be carefully used with the input frames.
While \cite{Wieschollek_2017_ICCV} designed a multi-scale architecture to deliver hierarchical information, 
\cite{Kim_2017_ICCV,Nah_2019_CVPR} modified the hidden states adaptively to the input in order to better focus on the target frame.
However, motion-blurred sequences often suffer from abrupt change of scenes that loosens the correlation between the adjacent frames.
Extending the neighborhood from the single previous frame to a predetermined range of past and future frames, \cite{zhong2020efficient} keeps a set of features and uses them to deblur the center frame.
Such an approach could look wider to find more relevant information from the saved features but requires extra RAM proportional to the size of the neighborhood.
Also, the optimal number of frames to look could vary by the local blur dynamics and scene contents.

Instead of saving multiple features or hidden states, we propose to model the long-range information of video in an extra memory with our recurrence-in-recurrence network~(RIRN).
On top of an RNN architecture, we design an inner-recurrence module~(IRM) that generates an auxiliary state with complementary information from the hidden state.
Different from \cite{zhong2020efficient}, we do not drop the oldest memory but let IRM learn to keep necessary information.
Also, we propose an adaptive temporal blending~(ATB) method that finds the relevant part of the information from the local neighborhood.
ATB generates attention maps on the image features from the current and the previous frames.
While we adopt the idea of temporal blending from \cite{Kim_2017_ICCV}, we relax the attention constraint so that features could be either attenuated or emphasized by the necessity in reconstructing the deblurred image via learning.

By conducting ablation study with a baseline method, we analyze the behavior of the proposed IRM and ATB and show the effectiveness of the architectural designs.
Furthermore, we apply the RIRN architecture to various RNN-based video deblurring methods as well as the gated architectures, LSTM~\cite{hochreiter1997long} and GRU~\cite{cho2014learning_gru} that manipulates the update of hidden states.
Our extensive experimental results exhibit the consistent improvements from our method both quantitatively and qualitatively.

\begin{figure}[t]
    \renewcommand{\wp}{0.16\linewidth}
    \captionsetup[subfloat]{font=scriptsize}
    \subfloat[GRU]{\includegraphics[width=\wp]{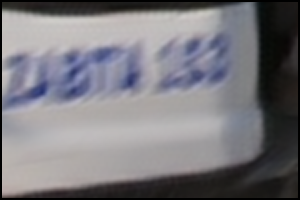}}
    \hfill
    \subfloat[LSTM]{\includegraphics[width=\wp]{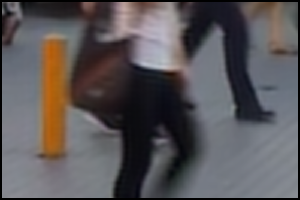}}
    \hfill
    \subfloat[STRCNN~\cite{Kim_2017_ICCV}]{\includegraphics[width=\wp]{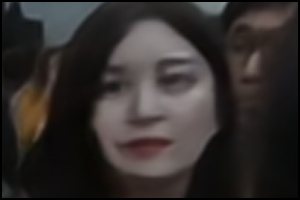}}
    \hfill
    \subfloat[STFAN~\cite{Zhou_2019_ICCV}]{\includegraphics[width=\wp]{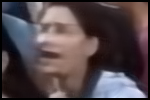}}
    \hfill
    \subfloat[IFI-RNN~\cite{Nah_2019_CVPR}]{\includegraphics[width=\wp]{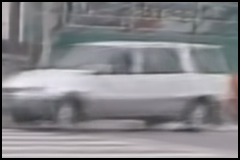}}
    \hfill
    \subfloat[RDBN~\cite{zhong2020efficient}]{\includegraphics[width=\wp]{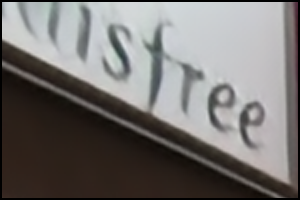}}
    \\
    \vspace{-3mm}
    \\
    \subfloat[GRU+RIRN]{\includegraphics[width=\wp]{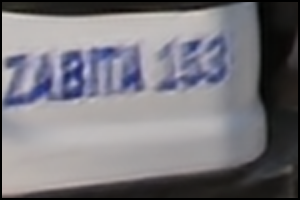}}
    \hfill
    \subfloat[LSTM+RIRN]{\includegraphics[width=\wp]{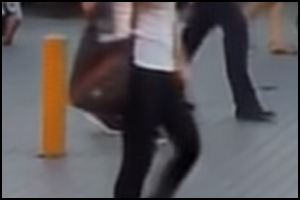}}
    \hfill
    \subfloat[STRCNN+RIRN]{\includegraphics[width=\wp]{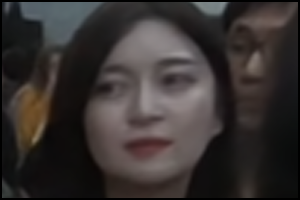}}
    \hfill
    \subfloat[STFAN+RIRN]{\includegraphics[width=\wp]{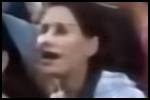}}
    \hfill
    \subfloat[IFI-RNN+RIRN]{\includegraphics[width=\wp]{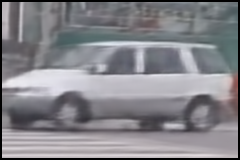}}
    \hfill
    \subfloat[RDBN+RIRN]{\includegraphics[width=\wp]{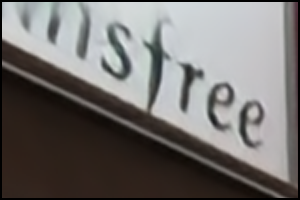}}
    \figcspace
    \caption{
        Visual comparison of RNN-based video deblurring results~(top) and our RIRN-applied results~(bottom).
        RIRN recovers the image details where the baseline methods fail.
    }
    \label{fig:overview}
    \figspace
\end{figure}

\section{Related Works}
\label{sec:related_works}


\noindent
\textbf{Video Deblurring with Explicit Motion Modeling.} 
While single image deblurring methods typically investigate the spatial information to handle the intrinsic blur trajectories~\cite{Nah_2017_CVPR,Tao_2018_CVPR,Zhang_2018_CVPR_svrn,ren2021deblurring,Gao_2019_CVPR}, early studies on video deblurring tried to aggregate the sharper information from the temporally neighboring frames~\cite{matsushita2006full,cho2012registration} by finding relevant patches.
To cope with complex blurry scenes, most of the optimization-based approaches estimated the blur kernel and used it to find the deblurred frames~\cite{li2010generating,Wulff_2014_ECCV,Kim_2015_CVPR,kim2017dynamic,Zhang_2014_CVPR} through the joint optimization process.
However, finding spatially non-uniform kernel for every pixel requires heavy computational resources.
In \cite{Su_2017_CVPR}, learning-based method was introduced by feeding a set of frames into CNN to deblur the center frame.
However, optical flow had to be computed from the blurry input frames to align them in the preprocessing.
Due to the blurriness, it is difficult to find the pixel-level correspondences and the error could cause inaccuracies in restoring the frames. 
In \cite{Kim_2018_ECCV}, spatio-temporal flow estimation was adopted to selectively capture the temporal dependencies while alleviating the occlusion problems from optical flow.
Later, \cite{Pan_2020_CVPR} proposed to improve the optical flow accuracy by estimating the flow and the latent frames simultaneously with neural networks using temporal sharpness prior.
In a recent work~\cite{li2021arvo}, the proposed ARVo method first uses an optical flow to align the frames.
Then, they construct a correlation volume pyramid among all pixel pairs in the neighboring frames to aggregate relevant information.

\noindent
\textbf{Video Deblurring with Implicit Motion Modeling.}
On the other hand, the motion information is implicitly employed in the approaches using recurrent neural networks.
Proposing a recurrent neural network architecture, RDN~\cite{Wieschollek_2017_ICCV} employed temporal skip connections in multiple feature scales to transfer information to the next frames.
In OVD~\cite{Kim_2017_ICCV}, the intermediate features were computed with dynamic temporal blending to improve the deblurring performance.
As such hidden states are generated from the past frames without knowing the target blurry frame, IFI-RNN used intra-frame iteration using the recurrence archtiecture to adapt the hidden states to better remove blur in the frame.
In a different approach, STFAN~\cite{Zhou_2019_ICCV} proposed a filter-adaptive network to align the features from multiple inputs within the neural network architecture from learning.
While \cite{Kim_2017_ICCV,Nah_2019_CVPR,Zhou_2019_ICCV} tried to better use the given hidden states generated from the previous time steps, ESTRNN~\cite{zhong2020efficient} proposed to use the features from both the past and the future frames.
From a set of features that are kept in memory, the spatio-temporal attention module fused the features to reconstruct the latent frame.
In recently proposed \cite{suin2021gated}, attention-based aggregation modules were proposed.
From the reinforcement learning-based keyframe batch selection, the gated spatio-temporal attention block uses non-local attention to gather sharper information.
By the attention-based aggregation process, the more useful information can be adaptively selected both spatially and temporally to better deblur a frame.
In contrast to \cite{zhong2020efficient} or \cite{suin2021gated}, our recurrence-in-recurrence architecture does not necessarily require future information or cached history of features.
We let our inner recurrence module to keep track of the temporal variance of hidden states from learning.

\noindent
\textbf{Long-Term Dependency Modeling in Sequential Data.}
LSTM~\cite{hochreiter1997long} is one of the earliest and widely-used architecture by solving vanishing gradient problem of RNNs.
The gates in LSTM determines the information to be forgotten in the hidden states.
GRU~\cite{cho2014learning_gru} presents an architecture with the reset and the update gates to manipulate hidden states.
While the gates are meant to model the long-term dependencies in sequential data, every information is saved in a single state which may eventually lead to short-term optimization~\cite{zhao2020rnn}.
Transformer architectures~\cite{vaswani2017attention} instead, stores the hidden activation at every time step and uses an attention module to integrate them, requiring large memory footprint.
To model longer-range dependencies in language modeling, Transformer-XL~\cite{dai2019transformer} reuses the states in segment-level recurrences, however, sufficiently old states are discarded determined by the size of memory.
Recently, \cite{rae2020compressive} introduced compressed representation to preserve old memories instead of discarding or saving in raw forms.
They extended the baseline transformer with an additional compression function.
Our recurrence-in-recurrence architecture shares a similar idea with \cite{rae2020compressive} in terms of augmenting the short-term memory from an auxiliary module generating additional hidden states.

\section{Proposed Method}
\label{sec:proposed_method}

\begin{figure}[t]
    \centering
    \renewcommand{\wp}{0.3\linewidth}
    \subfloat[Recurrence-in-Recurrence Network \label{fig:model_all}]{
    \includegraphics[width=0.9\linewidth]{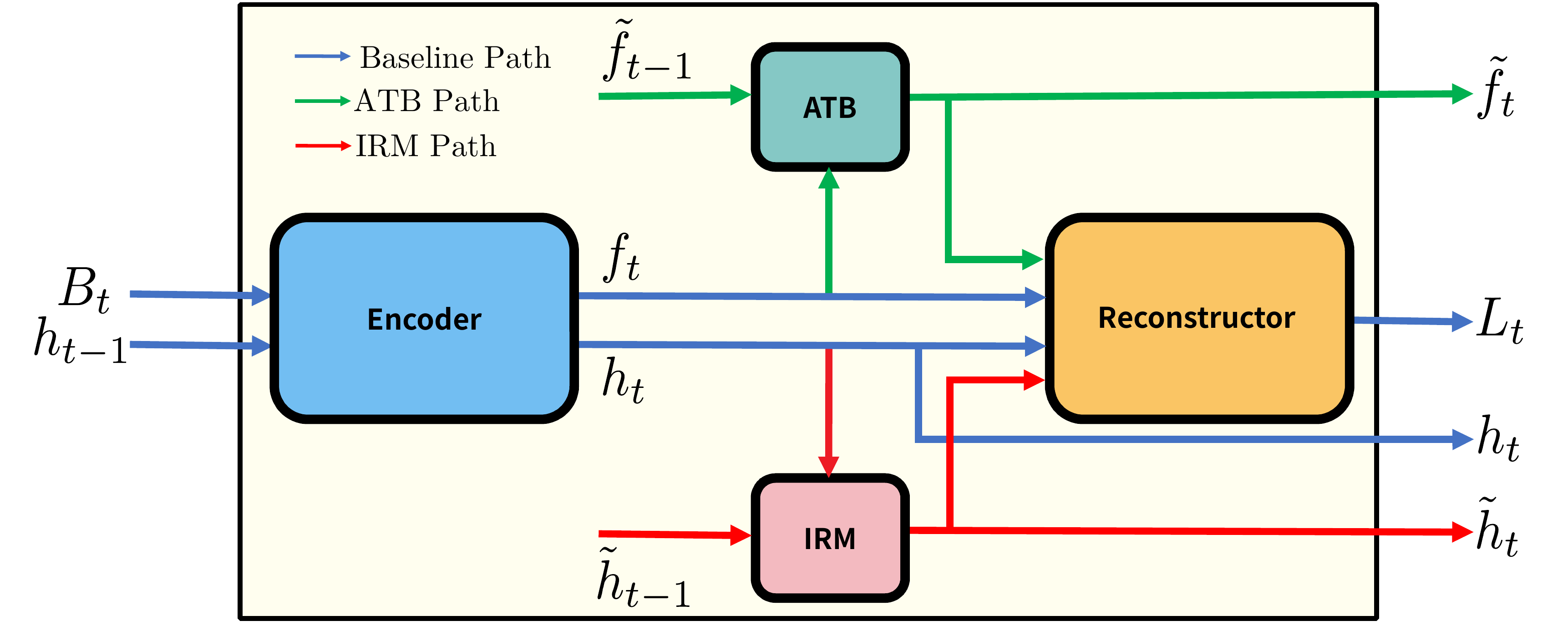}
    }
    \\
    \subfloat{\includegraphics[width=0.9\linewidth]{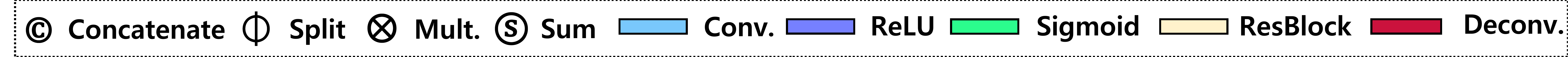}}
    \addtocounter{subfigure}{-1}
    \\
    \subfloat[Inner-Recurrence Module \label{fig:model_irm}]{
    \includegraphics[width=\wp]{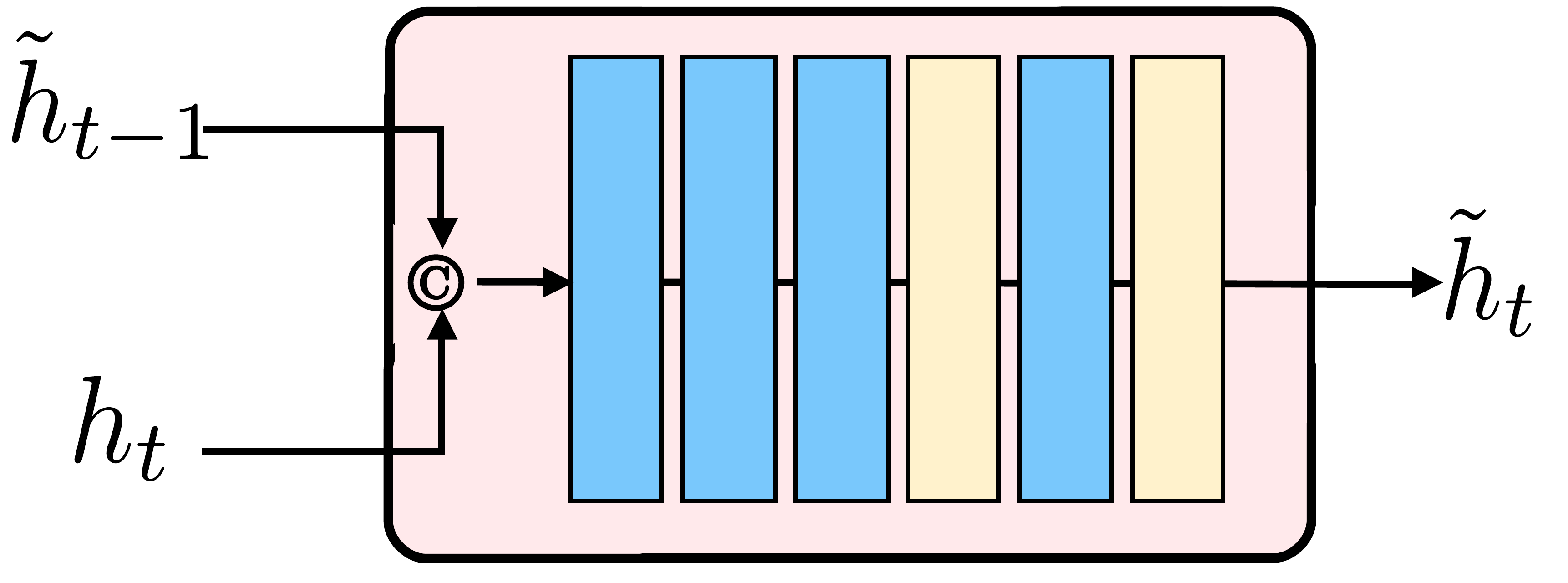}
    }
    \hfill
    \subfloat[Adaptive Temporal Blending \label{fig:model_atb}]{
    \includegraphics[width=\wp]{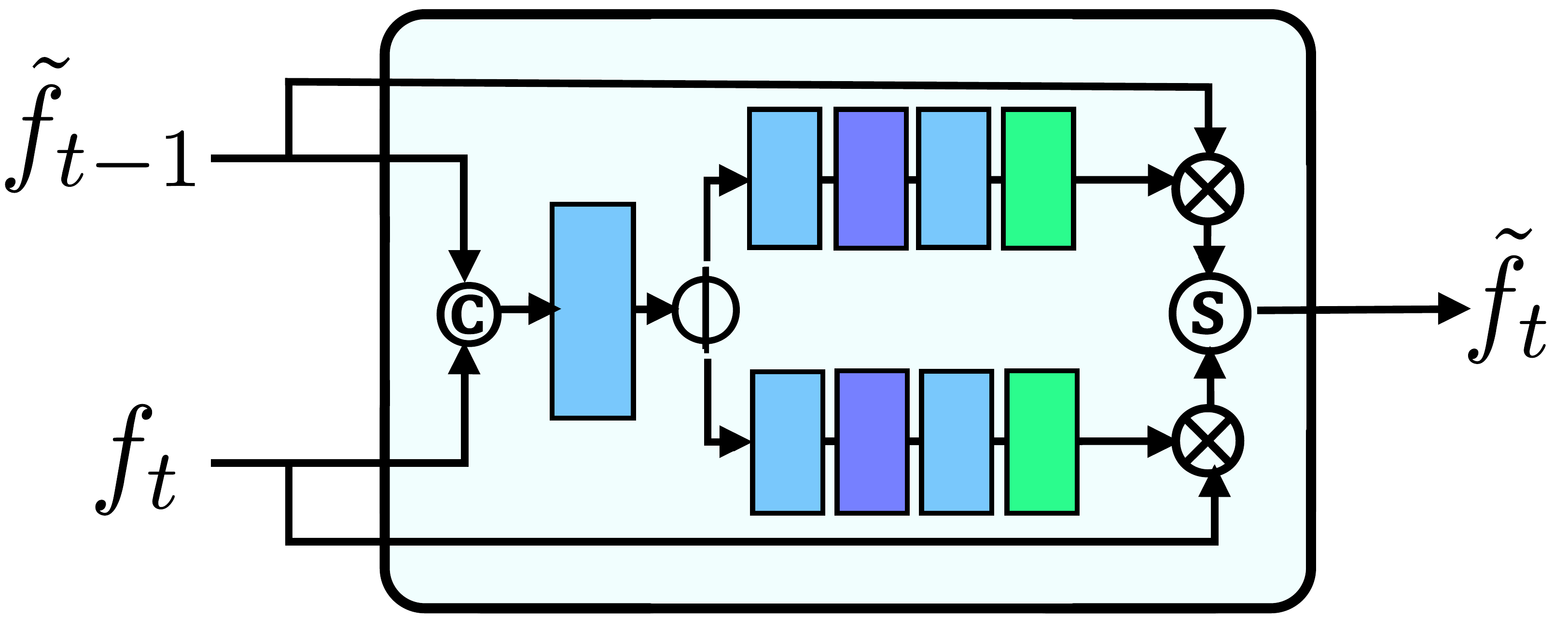}
    }
    \hfill
    \subfloat[Reconstructor Module \label{fig:model_recon}]{
    \includegraphics[width=\wp]{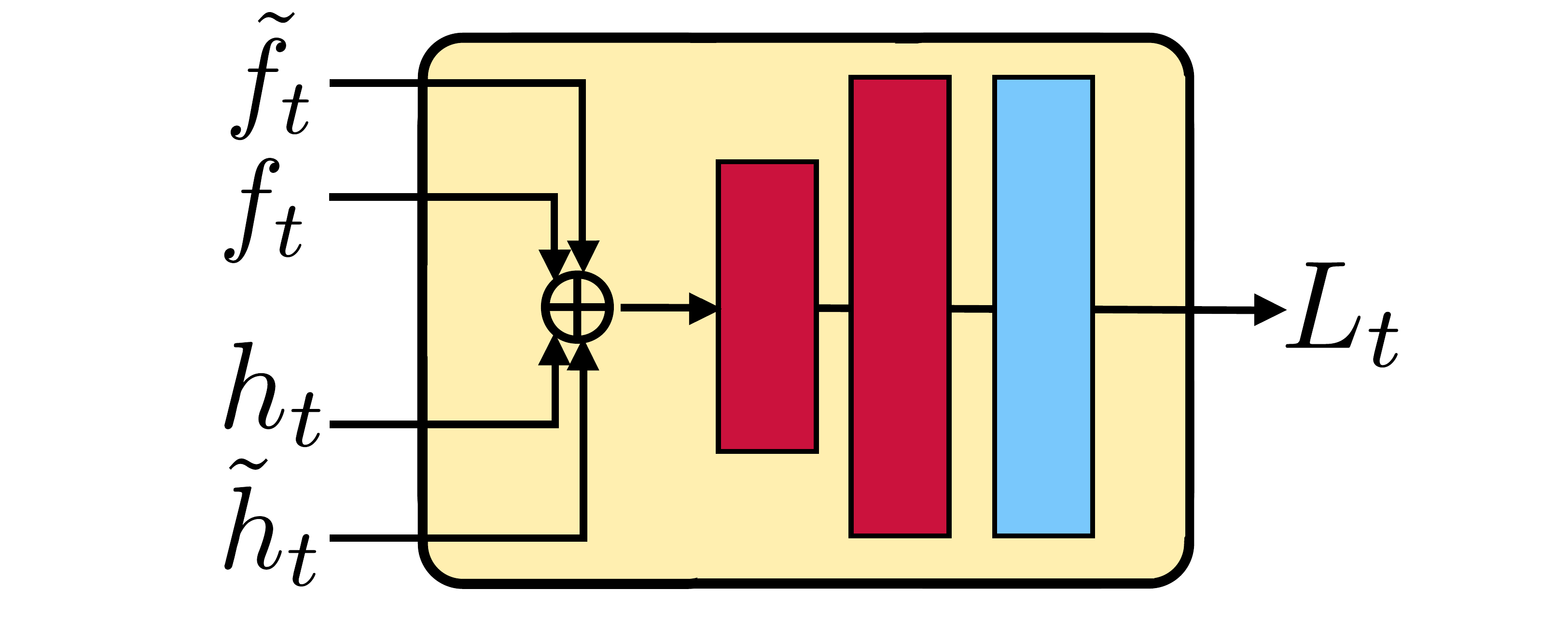}
    }
    \figcspace
    \caption{The architecture of Recurrence-in-Recurrence Network and the components}
    \label{fig:model}
\end{figure}

The main goal of this paper is to improve the video deblurring quality of RNN-based methods by supplementing the hidden states by providing an additional long-ranged memory.
Starting from the baseline RNN structure, we introduce an inner-recurrence module and the adaptive temporal blending scheme to better handle the temporal dependency.
The overall architecture is shown in Figure~\ref{fig:model}.


\subsection{Video Deblurring with Recurrent Networks}

We design a baseline recurrent network architecture with a feature extractor and a image reconstructor module.
From the input blurry frame $B_t$ and the hidden state $h_{t-1}$ from the previous time step, the feature encoder emits an updated hidden state $h_t$ and a feature $f_t$ as
\begin{equation}
    f_t, h_t = \text{Encoder}(B_t, h_{t-1}).
    \label{eq:encoder}
\end{equation}
Then, the outputs from the encoder is used to reconstruct the deblurred image $L_t$ as
\begin{equation}
    L_t = \text{Reconstructor}(f_t, h_t).
    \label{eq:recon}
\end{equation}
The recurrent model is trained with supervised loss function by comparing the output $L_t$ with the ground-truth sharp image $S_t$ as
$\norm{L_t - S_t}$.

In our proposed recurrence-in-recurrence network, the inner-recurrence module and adaptive temporal blending each supplement the hidden state $h_t$ and the feature ${f}_t$ to better deblur video frames.






\subsection{Inner-Recurrence Module}

{\revision{
RNNs basically use a single hidden state at each time step to store information from the past frames.
The hidden state at each step is optimized to maximize the deblurring performance of the corresponding frame.
At every time step, less relevant information to the target frame is forgotten and replaced by new information from the input.
However, dynamic videos tend to suffer from abrupt scene changes (i.e. camera shakes) so that long-ranged distant frames are often more related than direct neighbors.
In order to find and provide complementary information with long-term relation, we further exploit the recurrence operation to model the temporal variations of the hidden states.
Motivated that the set of hidden states, $\{h_{t}\}$, is another sequential data, we propose to design a new type of RNN whose inputs are the hidden states.
As the recurrence operates inside the standard architecture of RNNs, we term the module as Inner-Recurrence Module~(IRM) where 
\begin{equation}
    \tilde{h_{t}} = \text{IRM}(h_{t}, \tilde{h}_{t-1}).
    \label{eq:irm}
\end{equation}
The outputs of IRM, $\{\tilde{h}_{t}\}$ is generated by looking into the history and the temporal changes of $\{h_{t}\}$, serving as a sequence of higher-order memory states.
In contrast to saving multiple features in \cite{zhong2020efficient}, IRM does not require additional memory to store the cached set of states at inference.
Similarly to Compressive Transformer~\cite{rae2020compressive} storing the compressed memory from the memory sequence, our IRM preserves long-ranged information from learning.
In the following image reconstructor module, $\tilde{h_{t}}$ is jointly used with $h_{t}$ together, supplementing the deblurring performance.
The IRM architecture is shown in Figure~\ref{fig:model_irm}.

}}

\subsection{Adaptive Temporal Blending}
Besides using the hidden state $h_{t-1}$ to deliver information, the previous input frame $B_{t-1}$ is often used as well as $B_t$ in video deblurring RNNs~\cite{Zhou_2019_ICCV,zhong2020efficient} by concatenating the features.
Instead, we choose to adaptively blend the features with an attention-based recurrent module, inspired by dynamic temporal blending~\cite{Kim_2017_ICCV}.
Different from the IRM that manages the long-range dependency of frames, our Adaptive Temporal Blending~(ATB) focuses on the feature $f_t$ that is more specific to the target image at time $t$ as it is not propagated to the next frames in the baseline architecture.
{\revision{
DTB~\cite{Kim_2017_ICCV} assumes the features at all pixels to be equally important, forcing the 
sum-to-one constraints in the blending weights.
However, as the image feature is used together with the hidden states, there could be redundancy causing spatial variance in feature importance.
Thus, we adopted an attention mechanism so that the feature importance could be predicted from learning. 
}}
Instead of updating the input feature directly via recurrence as IRM, ATB produces the attention map to extract necessary information in $f_t$.
The spatial attention maps are multiplied to the features as conical combination,
\begin{equation}
    \tilde{f_{t}} = \tilde{w}_{t-1} \times \tilde{f}_{t-1} + w_t \times f_{t},
    \label{eq:atb}
\end{equation}
{\revision{
where $\tilde{w}_{t-1} \ge 0$ and $w_t \ge 0$.
Different from DTB, ATB does not require sum-to-one constraint.
}}
We let the attention weights to adaptively attenuate or emphasize the corresponding features without being tied to each other.
The design of ATB module is shown in Figure~\ref{fig:model_atb}.




\subsection{Reconstructor}

The reconstruction module aggregates all the features extracted from the encoder, inner-recurrence module, and the adaptive temporal blending module.
With all features $f_t$, $h_t$, $\tilde{f}_t$, $\tilde{h}_t$ concatenated, the deblurred image is obtained from several convolutional layers.
The reconstructor architecture is shown in Figure~\ref{fig:model_recon}.

\section{Experimental Results}

\subsection{Experimental Setup}
{\revision{To train and validate the performance of every model, we used GOPRO~\cite{Nah_2017_CVPR}, REDS~\cite{Nah_2019_CVPR_Workshops_REDS}, and DVD~\cite{Su_2017_CVPR} datasets.
DVD dataset contains 61 training sequences and 10 test sequences.
GOPRO dataset provides 22 training and 11 test sequences and REDS dataset has 240 training and 30 validation sequences.
All the video frames are in $1280\times720$ resolution.
To validate the effect of the proposed modules, every experiment was done in a unified setting, training from scratch with L1 loss and ADAM~\cite{kingma2014adam} optimizer with batch size 16. 
On GOPRO and REDS datasets, each model was trained for 500 and 200 epochs, respectively, annealing the learning rate at 300th and 100th epochs from the initial learning rate $1\times10^{-4}$.
On DVD dataset, models were trained for 500 epochs and the learning rate was halved after every 200 epochs.
For the model architectures used in the experiments and the implementation details, please refer to the supplementary material.}}

\subsection{Ablation Study: Effect of IRM and ATB}

{\revision{We validate the effect of our proposed IRM and ATB by applying them to the baseline RNN architecture based on RDBN~\cite{zhong2020efficient} and IFI-RNN~\cite{Nah_2019_CVPR}.}}
The simplest form of IFI-RNN without additional iterations, C1H1 is used.

\begin{table}[t]
    \captionsetup{font=footnotesize}
    \begin{minipage}[t]{0.46 \linewidth}
        \caption{Effect of ATB applied to RDBN}
        \tabcspace
        \label{tab:irm}
        \resizebox{\linewidth}{!}{
            \begin{tabular}{l|cc|cc|c}
                \toprule
                \multirow{2}{*}{Architecture} & \multicolumn{2}{c|}{GOPRO} & \multicolumn{2}{c|}{REDS} & \\
                 & PSNR & SSIM & PSNR & SSIM & time(sec)\\
                \midrule
                RDBN~\cite{zhong2020efficient} & 29.82 & 0.9043 & 32.29 & 0.9222 & 0.095\\
                RDBN + GSA & 30.10 & 0.9064 & 32.52 & 0.9233 & 0.170\\
                RDBN + IRM & \textbf{30.14} & \textbf{0.9072} & \textbf{32.59} & \textbf{0.9304} & 0.161\\
                \bottomrule
            \end{tabular}
        }
    \end{minipage}
    \hfill
    \begin{minipage}[t]{0.49 \linewidth}
        \caption{Effect of IRM applied to IFI-RNN (C1H1)}
        \tabcspace
        \label{tab:atb}
        \resizebox{\linewidth}{!}{
            \begin{tabular}{l|cc|cc|c}
                \toprule
                \multirow{2}{*}{Architecture} & \multicolumn{2}{c|}{GOPRO} & \multicolumn{2}{c|}{REDS} & \\
                 & PSNR & SSIM & PSNR & SSIM & time(sec)\\
                \midrule
                IFI-RNN~\cite{Nah_2019_CVPR} & 28.30 & 0.8668 & 30.01 & 0.8762 & 0.049\\
                IFI-RNN + DTB~\cite{Kim_2017_ICCV} & 28.31 & 0.8697 & 30.12 & 0.8763 & 0.059\\
                IFI-RNN + ATB & \textbf{28.65} & \textbf{0.8779} & \textbf{30.61} & \textbf{0.8800} & 0.065\\
                \bottomrule
            \end{tabular}
            }
    \end{minipage}
    \tabspace
\end{table}

\begin{figure}[t]
    \centering
    \renewcommand{\wp}{0.245\linewidth}
    \subfloat[${h}_{t-1}$ \label{fig:irn1_a}]{\includegraphics[width=\wp]{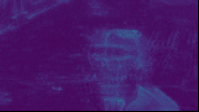}}
    \hfill
    \subfloat[$\tilde{h}_{t-1}$ \label{fig:irn1_b}]{\includegraphics[width=\wp]{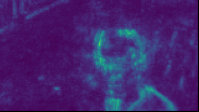}}
    \hfill
    \subfloat[$L_{t-1} - S_{t-1}$ \label{fig:irn1_c}]{\includegraphics[width=\wp]{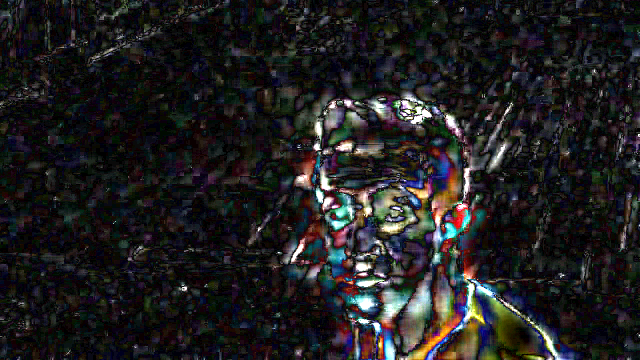}}
    \hfill
    \subfloat[${L}_{t}$ \label{fig:irn1_d}]{\includegraphics[width=\wp]{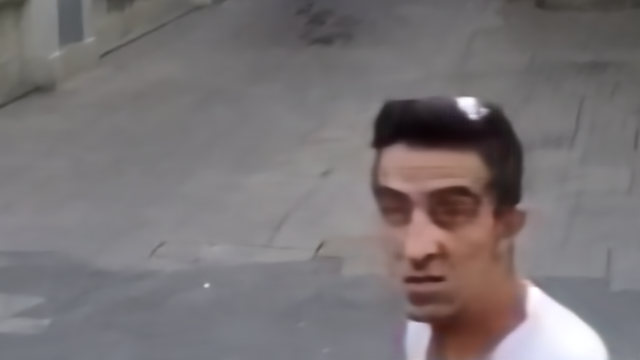}}
    \hfill
    \\
    \figcspace
    \caption{Visualization of hidden state and the IRM output}
    \label{fig:irn1}
    \figspace
\end{figure}

In Table~\ref{tab:irm}, we show the effect of applying IRM on the baseline model on GOPRO and REDS datasets.
Our IRM improves the deblurring accuracy by large margin at a similar degree of computing time.
We compare the effect of IRM and GSA~\cite{zhong2020efficient} by saving a set of features from different time steps.
As we do not use future frames for deblurred frame estimation, we used 3 features, $\{f_{t-2}, f_{t-1}, f_{t}\}$, before updating the hidden state $h_{t}$. 
We find that IRM is not only more accurate than GSA but also faster in computation.

In Figure~\ref{fig:irn1}, we visualize the hidden state and the long-term memory obtained from the IRM as well as the error map of the deblurred image from the previous time step.
While the facial texture requires detailed information, the hidden state $h_{t-1}$ fails to serve as an informative cue due to the erroneous estimation in the previous frame as shown in Figure~\ref{fig:irn1_c}.
{\revision{It shows that}} $\tilde{h}_{t-1}$ from IRM brings a complementary information from the further past frames to help recover $L_t$.

\begin{figure}[t]
    \renewcommand{\wp}{0.49\linewidth}
    \subfloat[IFIC1 + ATB \label{fig:distribution_a}]{\includegraphics[width=\wp]{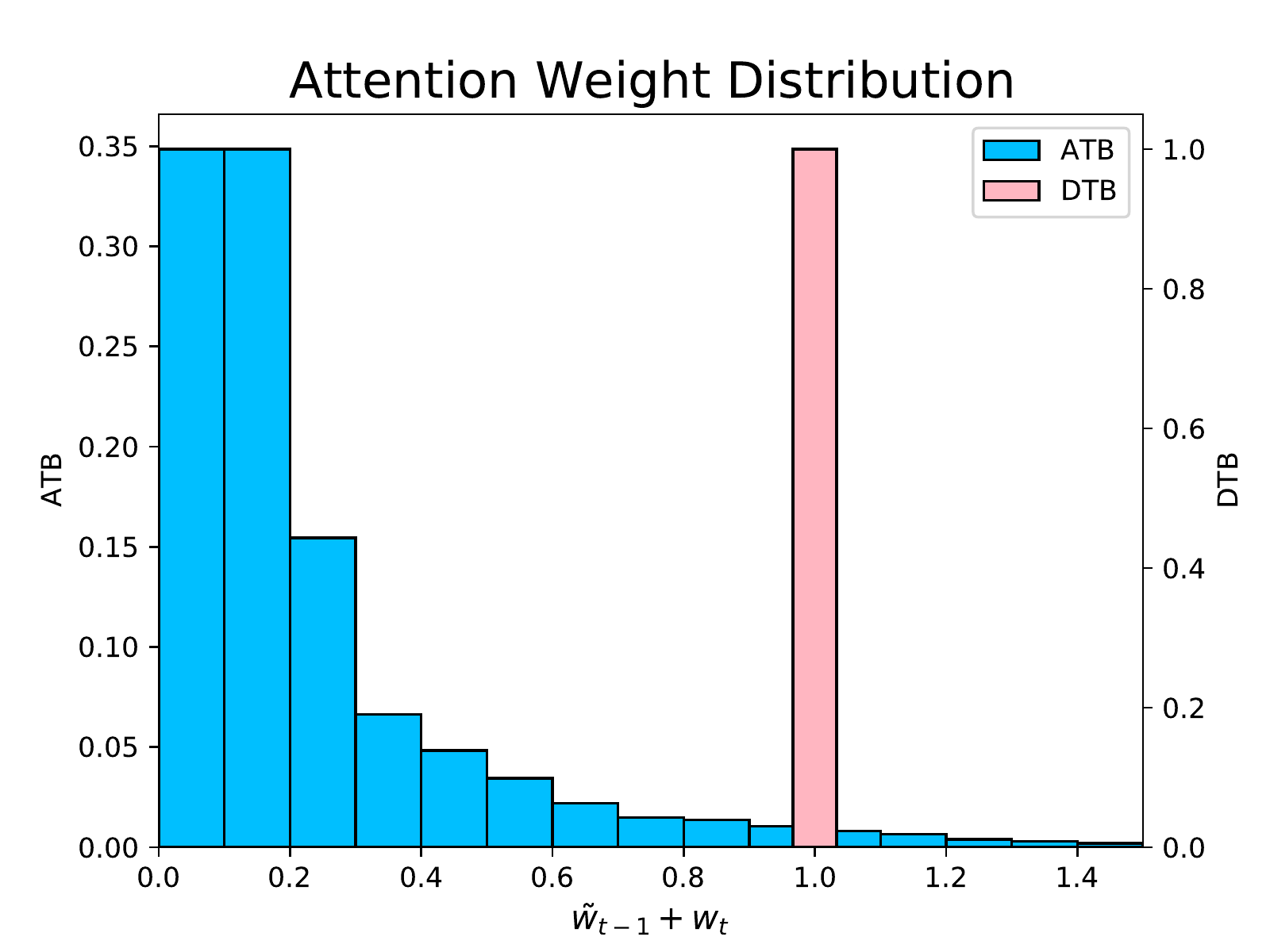}}
    \hfill
    \subfloat[IFIC1 + IRM + ATB \label{fig:distribution_b}]{\includegraphics[width=\wp]{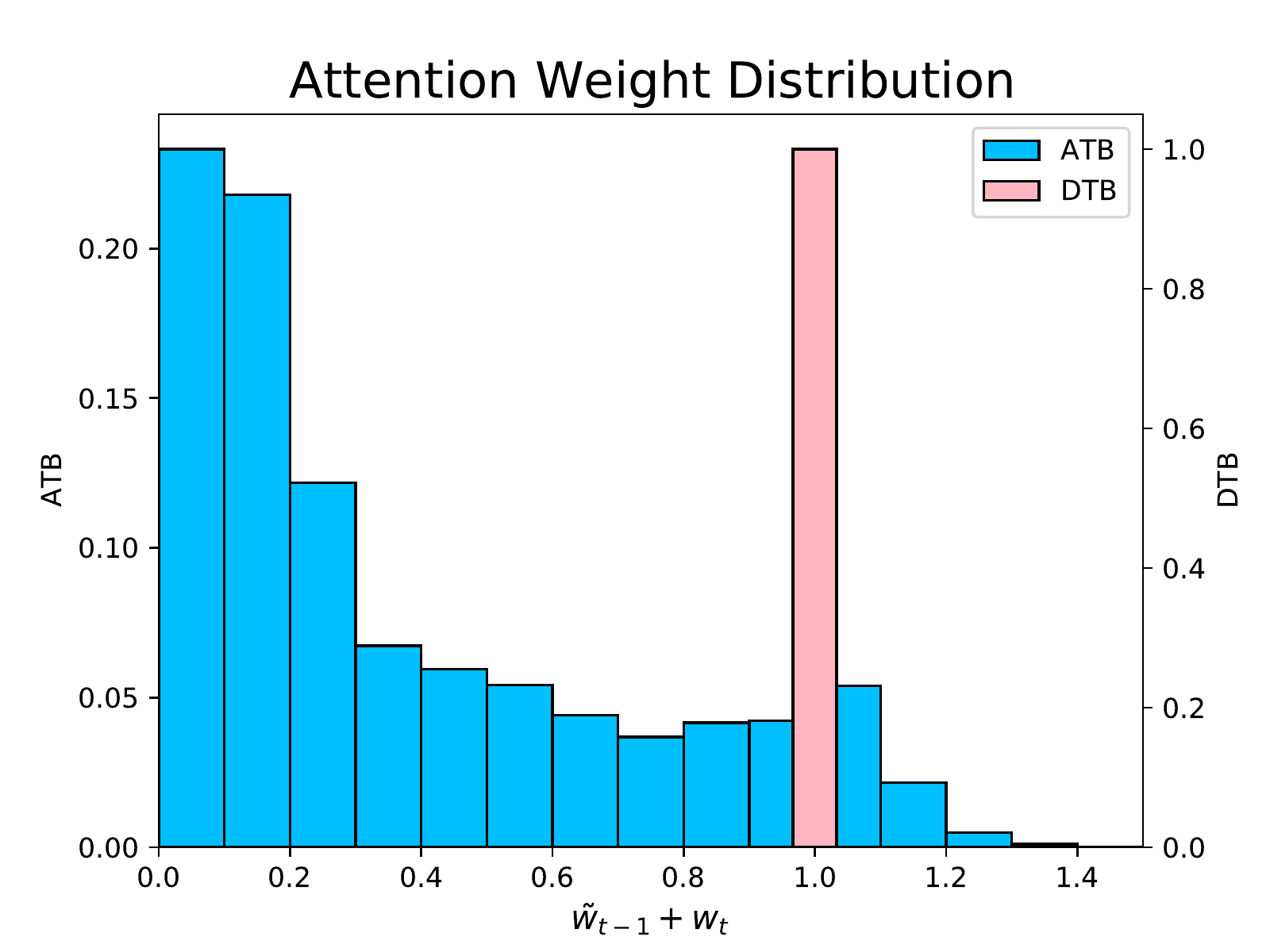}}
    \figcspace
    \caption{Distribution of sum of attention weights in ATB and DTB on GOPRO~\cite{Nah_2017_CVPR} dataset.}
    \label{fig:disribution}
    \figspace
\end{figure}
\begin{figure}[t]
    \centering
    \subfloat[$B_t$ \label{fig:atb_a}]{\includegraphics[width=0.13\linewidth]{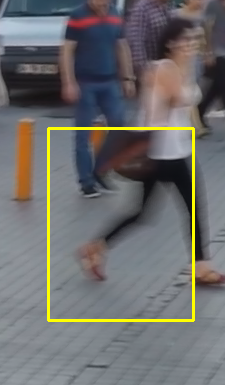}}
    \hspace{0.1cm}
    \subfloat[$f_{t}$ \label{fig:atb_b}]{\includegraphics[width=0.13\linewidth]{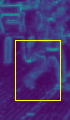}}
    \hspace{0.1cm}
    \subfloat[DTB~\cite{Kim_2017_ICCV} \label{fig:atb_c}]{\includegraphics[width=0.13\linewidth]{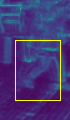}}
    \hspace{0.1cm}
    \subfloat[ATB \label{fig:atb_d}]{\includegraphics[width=0.13\linewidth]{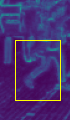}}
    \hspace{0.1cm}
    \subfloat[$L_{t}$ \label{fig:atb_e}]{\includegraphics[width=0.13\linewidth]{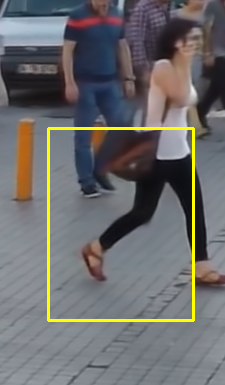}}
    \hspace{0.1cm}
    \\
    \figcspace
    \caption{Visualization of $\Tilde{f}_{t}$ in baseline model(IFI-RNN) with using DTB and ATB.}
    \vspace{-3mm}
    \label{fig:atb}
\end{figure}

In Table~\ref{tab:atb}, we validate the effect of blending features by using ATB in the RNN architecture.
Our ATB successfully improves the deblurring performance of the baseline while DTB provides marginal gains over the baseline.
The different behavior of ATB and DTB in handling the features is shown in Figure~\ref{fig:disribution}.
By removing the sum-to-one constraint of DTB in our adaptive temporal blending, we find that majority of the features are suppressed by the learned attention.
It shows that our ATB could better find the useful constituents in the features while discarding the unnecessary information.
Interestingly, when IRM is used jointly, ATB tends to makes more use of the $f_t$ and $\tilde{f}_{t-1}$ as shown in Figure~\ref{fig:distribution_b}.
{\revision{
We further visualize the effectiveness of our ATB in Figure~\ref{fig:atb} by showing the blended feature.
While DTB fails to gather necessary information from blending in Figure~\ref{fig:atb_c}, ATB shows clearer edges in Figure~\ref{fig:atb_d}, helping the reconstruction from the walking person's blurry legs.
It indicates that our relaxed condition from removing the sum-to-one constraint leads to successful selection on necessary information from the spatial attention.

}}

\begin{figure}[t]
    \renewcommand{\wp}{0.49\linewidth}
    \newcommand{\wwp}{0.35\linewidth}
    \newcommand{\hp}{\hspace{1mm}}
    \captionsetup[subfloat]{font=scriptsize}
    
    \begin{minipage}{0.6\textwidth}
        \centering
        \subfloat[Blur]{\includegraphics[width=\wp]{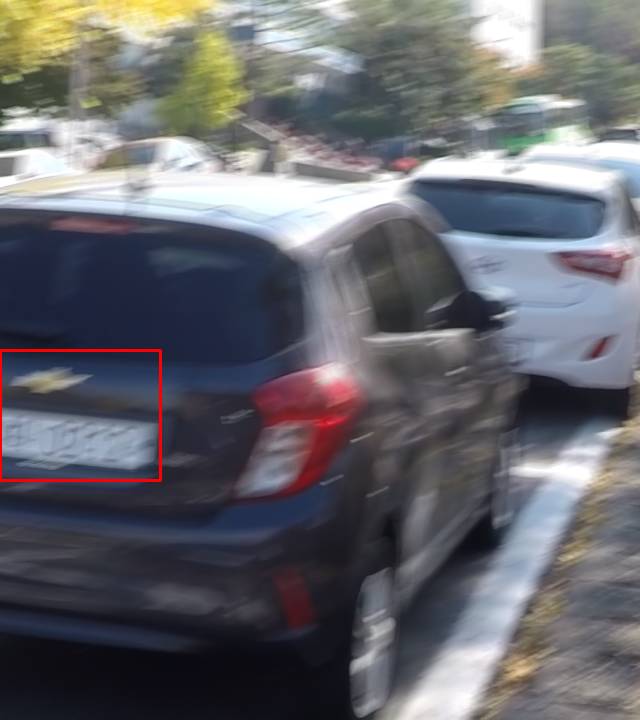}}
        \hp
        \subfloat[Deblurred (IFIC1+ATB+IRM)]{\includegraphics[width=\wp]{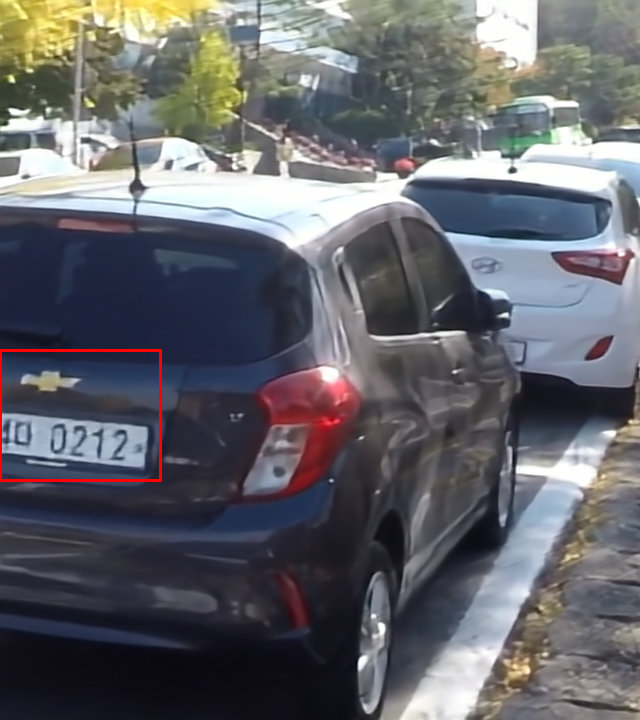}}
        \hp
    \end{minipage}
    \begin{minipage}{0.5\textwidth}
        \subfloat[IFIC1~\cite{Nah_2019_CVPR}]{\includegraphics[width=\wwp]{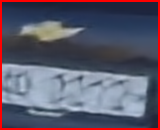}}
        \hp
        \subfloat[IFIC1+ATB]{\includegraphics[width=\wwp]{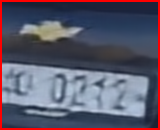}}
        \\
        \vspace{-2mm}
        \\
        \subfloat[IFIC1+IRM]{\includegraphics[width=\wwp]{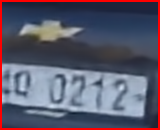}}
        \hp
        \subfloat[IFIC1+ATB+IRM]{\includegraphics[width=\wwp]{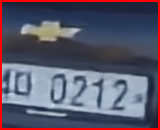}}
    \end{minipage}
    \figcspace
    \caption{Visual ablation showing the effect of ATB and IRM on GOPRO~\cite{Nah_2017_CVPR} dataset.}
    \label{fig:ablation}
\end{figure}

In Figure~\ref{fig:ablation}, we visualize the effect of IRM, ATB in the deblurred results.
The carplate numbers are better recognizable from the results using both modules.

\subsection{Effect of RIRN in State-of-The-Art Methods}

\begin{figure}[t]
    \captionsetup[subfloat]{font=footnotesize}
    \newcommand{\ww}{0.16\linewidth}
    \centering
    \subfloat[Input blurry image]{\includegraphics[width=0.49\linewidth]{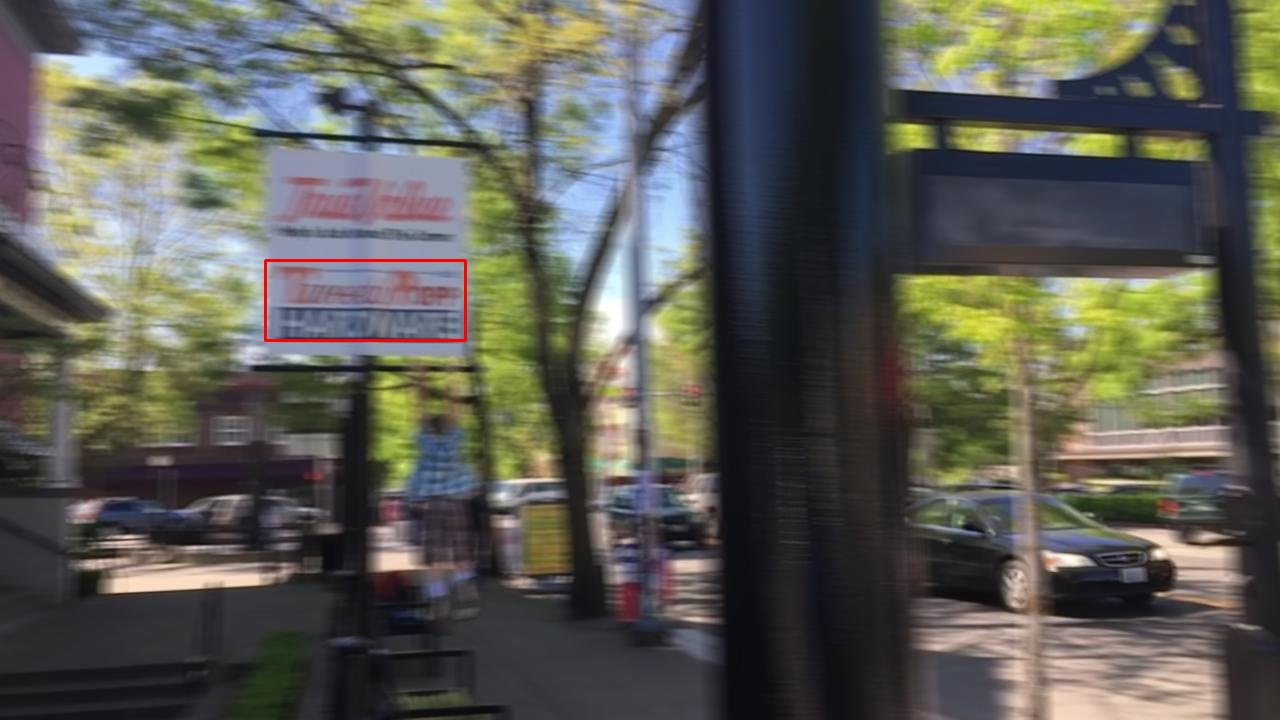}}
    \hfill
    \subfloat[Our deblurred image, RDBN+RIRN]{\includegraphics[width=0.49\linewidth]{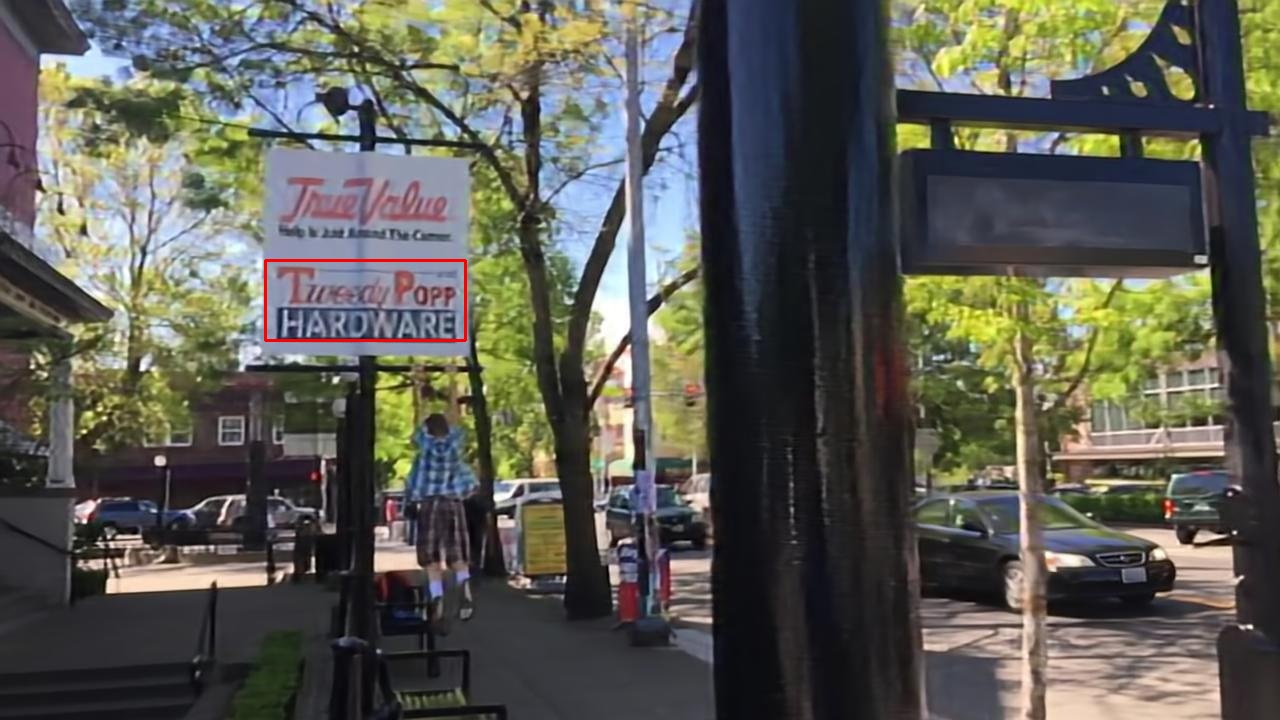}}
    \\
    \captionsetup[subfloat]{font=scriptsize}
    \vspace{-3mm}
    \subfloat[GRU]{\includegraphics[width=\ww]{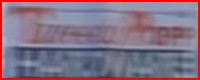}}
    \hfill
    \subfloat[LSTM]{\includegraphics[width=\ww]{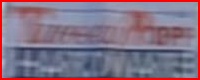}}
    \hfill
    \subfloat[STRCNN~\cite{Kim_2017_ICCV}]{\includegraphics[width=\ww]{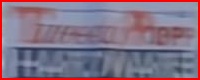}}
    \hfill
    \subfloat[STFAN~\cite{Zhou_2019_ICCV}]{\includegraphics[width=\ww]{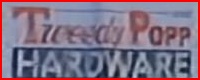}}
    \hfill
    \subfloat[IFIC1~\cite{Nah_2019_CVPR}]{\includegraphics[width=\ww]{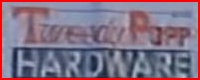}}
    \hfill
    \subfloat[RDBN~\cite{zhong2020efficient}]{\includegraphics[width=\ww]{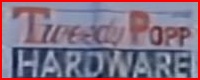}}
    \\
    \vspace{-3mm}
    \subfloat[GRU+RIRN]{\includegraphics[width=\ww]{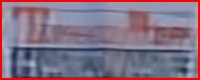}}
    \hfill
    \subfloat[LSTM+RIRN]{\includegraphics[width=\ww]{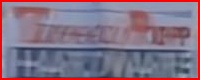}}
    \hfill
    \subfloat[STRCNN+RIRN]{\includegraphics[width=\ww]{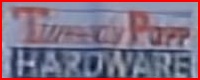}}
    \hfill
    \subfloat[STFAN+RIRN]{\includegraphics[width=\ww]{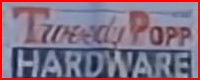}}
    \hfill
    \subfloat[IFIC1+RIRN]{\includegraphics[width=\ww]{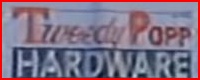}}
    \hfill
    \subfloat[RDBN+RIRN]{\includegraphics[width=\ww]{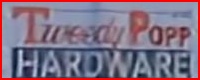}}
    \figcspace
    \caption{Comparison of baselines and RIRN-applied method results in DVD~\cite{Su_2017_CVPR} dataset}
    \label{fig:dvd_comparison}
    \figspace
\end{figure}

\begin{figure}[h]
    \captionsetup[subfloat]{font=footnotesize}
    \newcommand{\ww}{0.16\linewidth}
    \centering
    \subfloat[Input blurry image]{\includegraphics[width=0.49\linewidth]{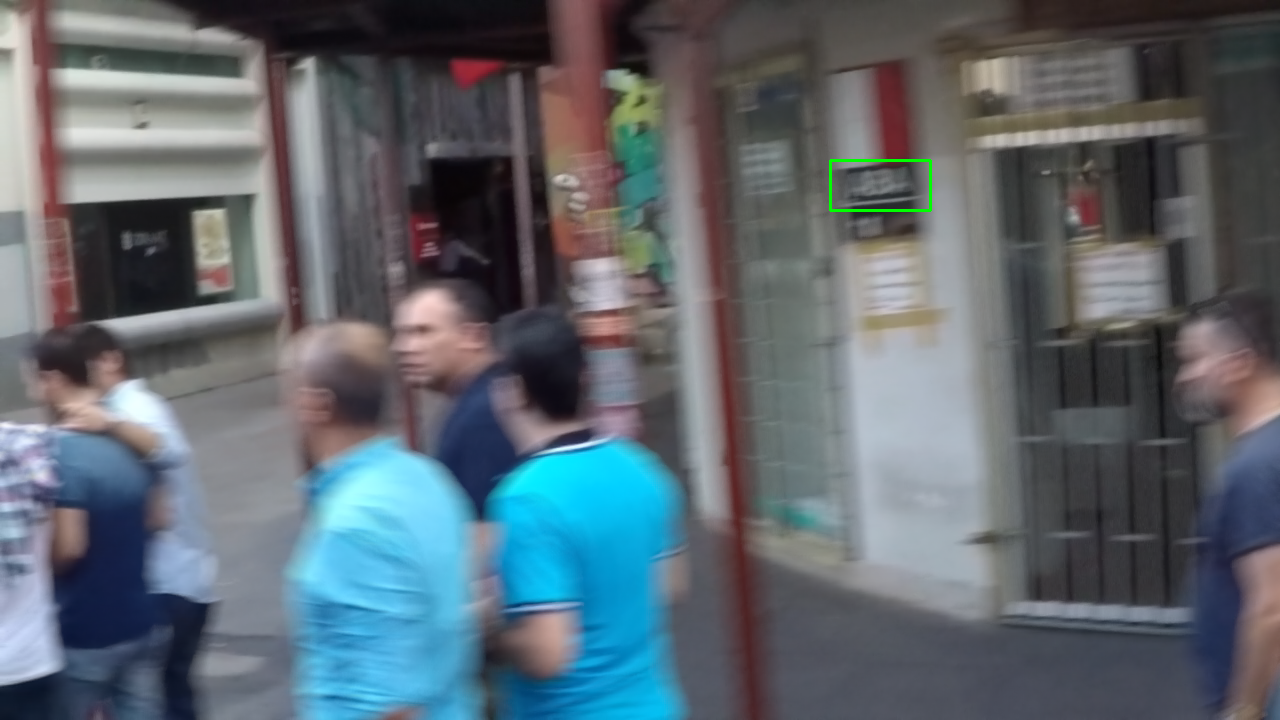}}
    \hfill
    \subfloat[Deblurred image (RDBN+RIRN)]{\includegraphics[width=0.49\linewidth]{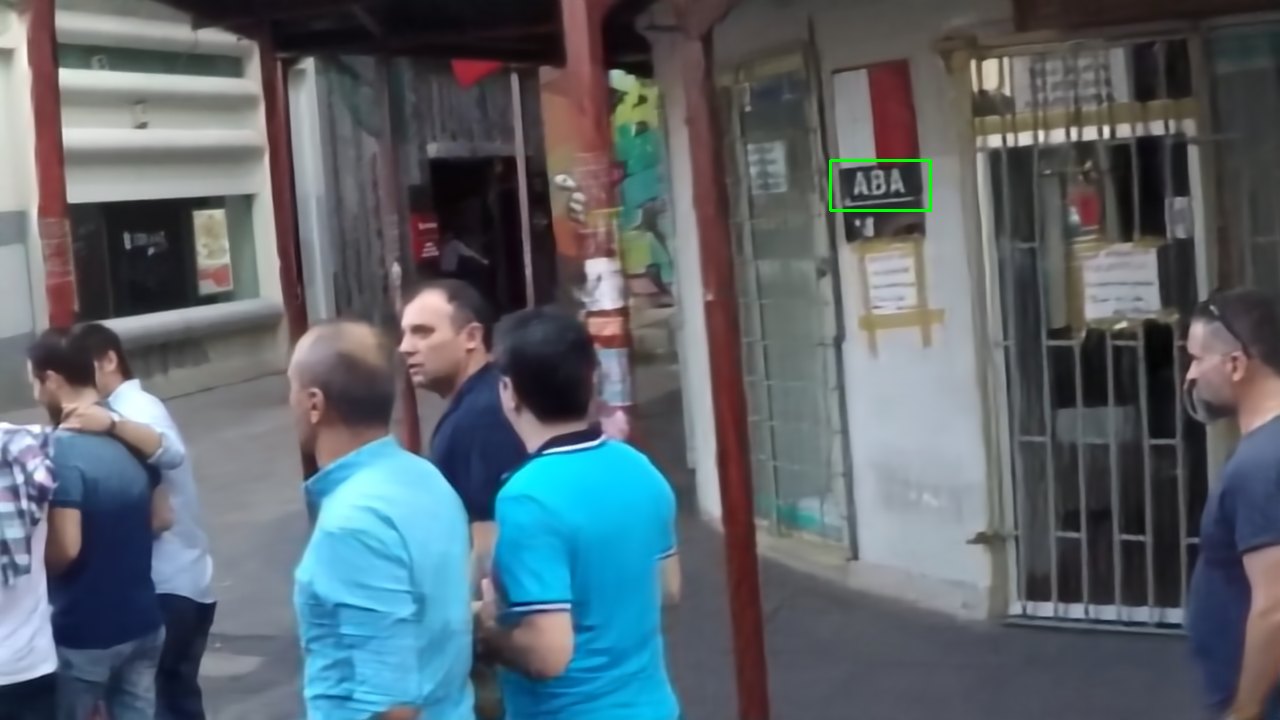}}
    \\
    \captionsetup[subfloat]{font=scriptsize}
    \subfloat[GRU]{\includegraphics[width=\ww]{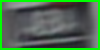}}
    \hfill
    \subfloat[LSTM]{\includegraphics[width=\ww]{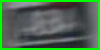}}
    \hfill
    \subfloat[STRCNN~\cite{Kim_2017_ICCV}]{\includegraphics[width=\ww]{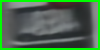}}
    \hfill
    \subfloat[STFAN~\cite{Zhou_2019_ICCV}]{\includegraphics[width=\ww]{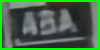}}
    \hfill
    \subfloat[IFIC1~\cite{Nah_2019_CVPR}]{\includegraphics[width=\ww]{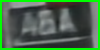}}
    \hfill
    \subfloat[RDBN~\cite{zhong2020efficient}]{\includegraphics[width=\ww]{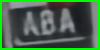}}
    \\
    \vspace{-2mm}
    \subfloat[GRU+RIRN]{\includegraphics[width=\ww]{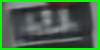}}
    \hfill
    \subfloat[LSTM+RIRN]{\includegraphics[width=\ww]{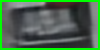}}
    \hfill
    \subfloat[STRCNN+RIRN]{\includegraphics[width=\ww]{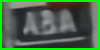}}
    \hfill
    \subfloat[STFAN+RIRN]{\includegraphics[width=\ww]{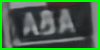}}
    \hfill
    \subfloat[IFIC1+RIRN]{\includegraphics[width=\ww]{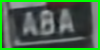}}
    \hfill
    \subfloat[RDBN+RIRN]{\includegraphics[width=\ww]{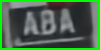}}
    \figcspace
    \caption{Comparison of the baselines and the RIRN-applied results on GOPRO~\cite{Nah_2017_CVPR} dataset}
    \label{fig:gopro_comparison}
    \vspace{-5mm}
\end{figure}

\begin{figure}
    \captionsetup[subfloat]{font=footnotesize}
    \newcommand{\ww}{0.16\linewidth}
    \centering
    \subfloat[Input blurry image]{\includegraphics[width=0.49\linewidth]{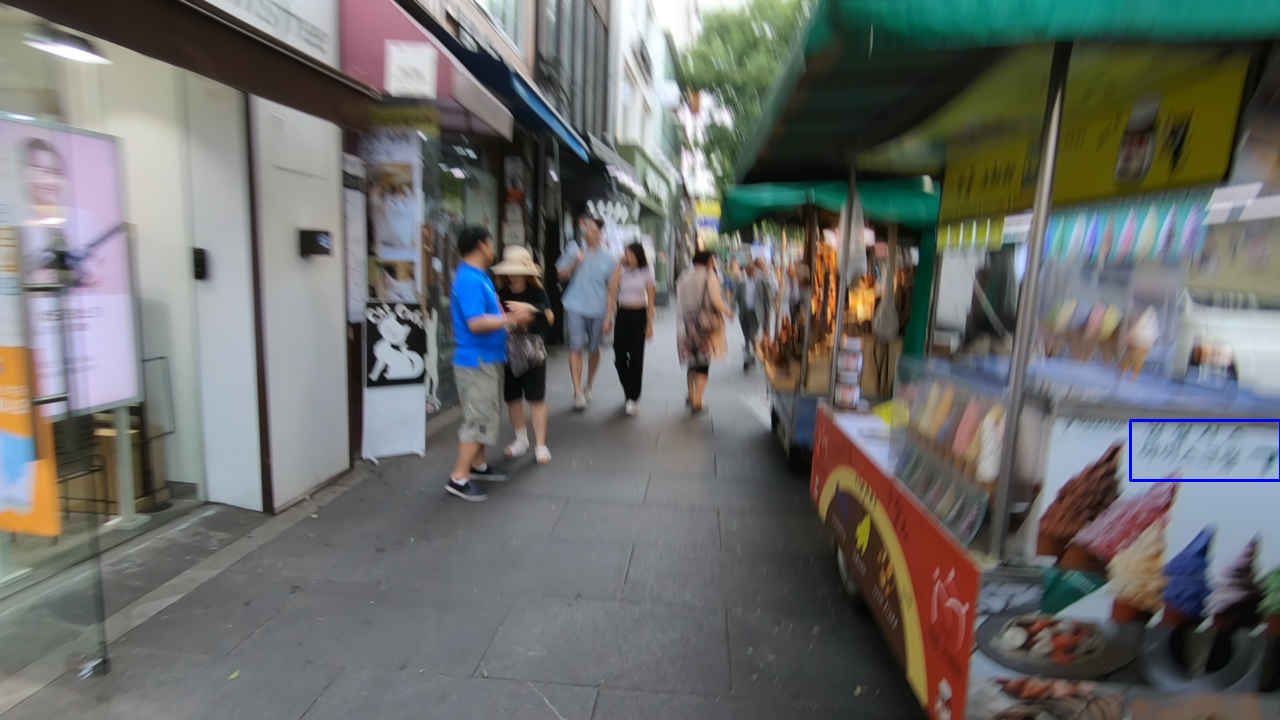}}
    \hfill
    \subfloat[Deblurred image (RDBN+RIRN)]{\includegraphics[width=0.49\linewidth]{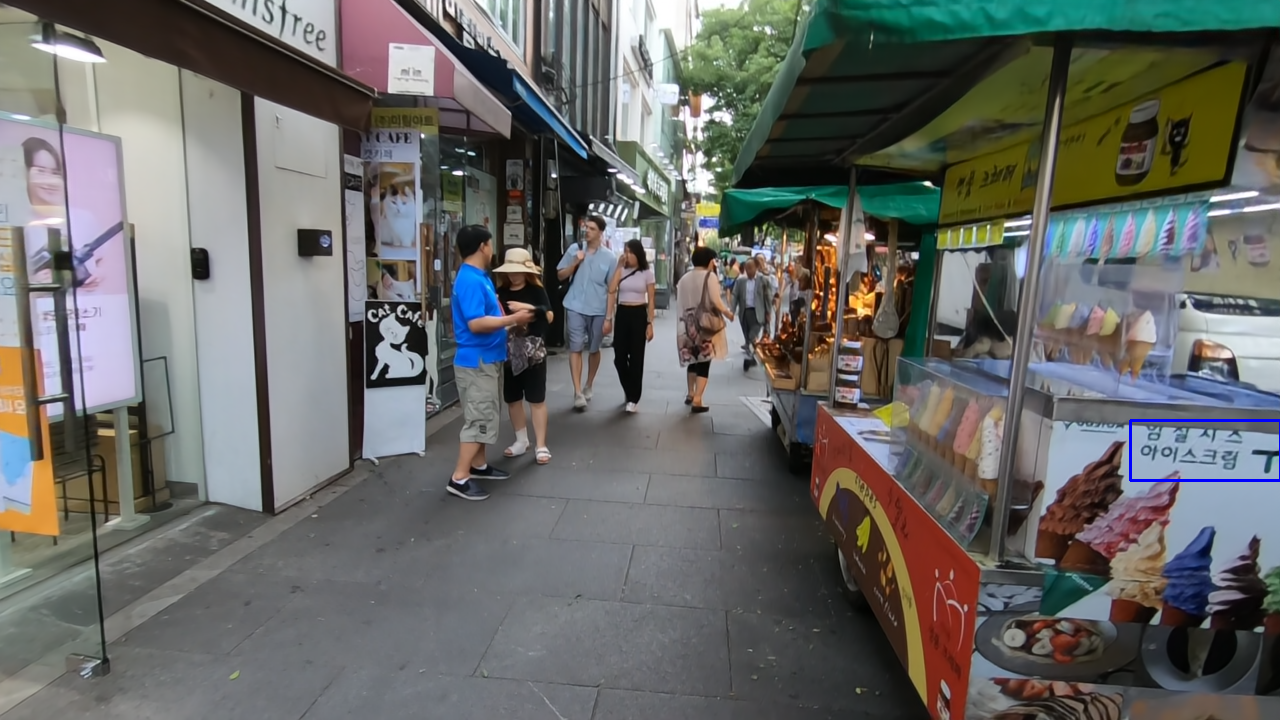}}
    \\
    \captionsetup[subfloat]{font=scriptsize}
    \vspace{-3mm}
    \subfloat[GRU]{\includegraphics[width=\ww]{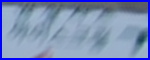}}
    \hfill
    \subfloat[LSTM]{\includegraphics[width=\ww]{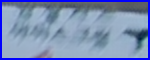}}
    \hfill
    \subfloat[STRCNN~\cite{Kim_2017_ICCV}]{\includegraphics[width=\ww]{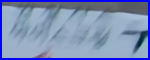}}
    \hfill
    \subfloat[STFAN~\cite{Zhou_2019_ICCV}]{\includegraphics[width=\ww]{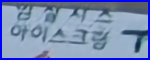}}
    \hfill
    \subfloat[IFIC1~\cite{Nah_2019_CVPR}]{\includegraphics[width=\ww]{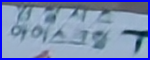}}
    \hfill
    \subfloat[RDBN~\cite{zhong2020efficient}]{\includegraphics[width=\ww]{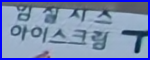}}
    \\
    \vspace{-3mm}
    \subfloat[GRU+RIRN]{\includegraphics[width=\ww]{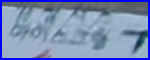}}
    \hfill
    \subfloat[LSTM+RIRN]{\includegraphics[width=\ww]{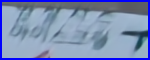}}
    \hfill
    \subfloat[STRCNN+RIRN]{\includegraphics[width=\ww]{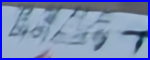}}
    \hfill
    \subfloat[STFAN+RIRN]{\includegraphics[width=\ww]{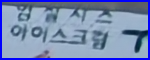}}
    \hfill
    \subfloat[IFIC1+RIRN]{\includegraphics[width=\ww]{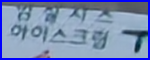}}
    \hfill
    \subfloat[RDBN+RIRN]{\includegraphics[width=\ww]{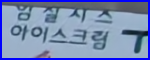}}
    \figcspace
    \caption{Comparison of the baselines and the RIRN-applied results on REDS~\cite{Nah_2019_CVPR_Workshops_REDS} dataset}
    \label{fig:reds_comparison}
    \vspace{-5mm}
\end{figure}

\begin{figure}
    \renewcommand{\wp}{0.16\linewidth}
    \captionsetup[subfloat]{font=scriptsize}
    \subfloat[GRU]{\includegraphics[width=\wp]{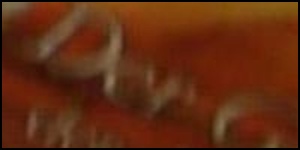}}
    \hfill
    \subfloat[LSTM]{\includegraphics[width=\wp]{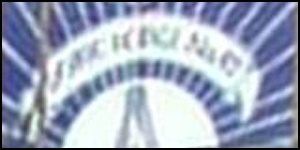}}
    \hfill
    \subfloat[STRCNN~\cite{Kim_2017_ICCV}]{\includegraphics[width=\wp]{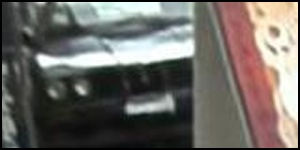}}
    \hfill
    \subfloat[STFAN~\cite{Zhou_2019_ICCV}]{\includegraphics[width=\wp]{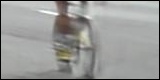}}
    \hfill
    \subfloat[IFI-RNN~\cite{Nah_2019_CVPR}]{\includegraphics[width=\wp]{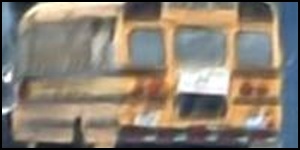}}
    \hfill
    \subfloat[RDBN~\cite{zhong2020efficient}]{\includegraphics[width=\wp]{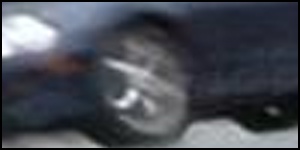}}
    \\
    \vspace{-3mm}
    \\
    \subfloat[GRU+RIRN]{\includegraphics[width=\wp]{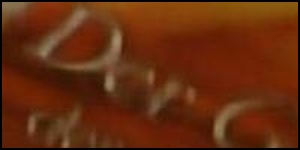}}
    \hfill
    \subfloat[LSTM+RIRN]{\includegraphics[width=\wp]{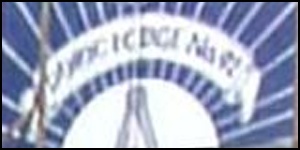}}
    \hfill
    \subfloat[STRCNN+RIRN]{\includegraphics[width=\wp]{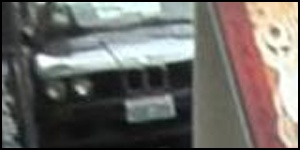}}
    \hfill
    \subfloat[STFAN+RIRN]{\includegraphics[width=\wp]{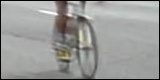}}
    \hfill
    \subfloat[IFI-RNN+RIRN]{\includegraphics[width=\wp]{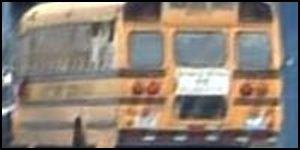}}
    \hfill
    \subfloat[RDBN+RIRN]{\includegraphics[width=\wp]{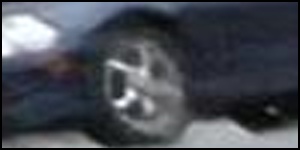}}
    \figcspace
    \caption{
        Visual comparison of RNN-based video deblurring results~(top) and our RIRN-applied results~(bottom) on real blurry videos.
    }
    \label{fig:real}
\end{figure}

Finding both the IRM and ATB to be effective in recurrent networks, we apply the recurrence-in-recurrence architecture to various video deblurring methods.
In Table~\ref{tab:sota_results}, we validate the generalizability of RIRN across different architectures by showing the consistent improvements from RIRN-applied models.
We use IFI-RNN~\cite{Nah_2019_CVPR} C1H1 model, STRCNN~\cite{Kim_2017_ICCV}, STFAN~\cite{Zhou_2019_ICCV}, RDBN.
RDBN is derived from ESTRNN~\cite{zhong2020efficient} but does not use the cached memory and GSA module.
Also, we apply our RIRN to the gated RNN architectures such as LSTM and GRU. where the updates of hidden states are controlled by the gates.
While the LSTM and GRU use the gates to control memory of RNNs, our RIRN successfully brings gains in deblurring accuracy from our IRN and ATB.
\begin{table}[t]
    \centering
    \caption{Application of RIRN on existing RNN architectures}
    \tabcspace
    \label{tab:sota_results}
    \footnotesize
    \begin{tabular}{c|cc|cc|cc|c}
        \toprule
        \multirow{2}{*}{Architecture} & \multicolumn{2}{c|}{DVD} & \multicolumn{2}{c|}{GOPRO}&\multicolumn{2}{c|}{REDS} \\
        & PSNR & SSIM & PSNR & SSIM & PSNR & SSIM & FPS\\
        \midrule
        IFI-RNN~\cite{Nah_2019_CVPR} & 30.53 & 0.9069 & 28.30 & 0.8668 & 30.01 & 0.8762 & 20.15\\
        IFI-RNN + RIRN & \textbf{30.97} & \textbf{0.9168} & \textbf{29.14} & \textbf{0.8894} & \textbf{31.08} & \textbf{0.8905} & 12.48\\
        \midrule
        STRCNN~\cite{Kim_2017_ICCV} & 29.15 & 0.8728 & 28.72 & 0.8460 & 30.23 & 0.8708 & 7.69\\
        STRCNN + RIRN & \textbf{30.17} & \textbf{0.9019} & \textbf{28.87} & \textbf{0.8781} & \textbf{30.76} & \textbf{0.8902} & 5.94\\
        \midrule
        STFAN~\cite{Zhou_2019_ICCV}& 30.93 & 0.9087 & 28.77 & 0.8776 & 31.26 & 0.8864 & 6.66\\
        STFAN + RIRN & \textbf{31.03} & \textbf{0.9096} & \textbf{29.24} & \textbf{0.8876} & \textbf{31.44} & \textbf{0.8951} & 4.42\\
        \midrule
        RDBN~\cite{zhong2020efficient}& 31.44 & 0.9188 & 29.82 & 0.9043 & 32.29 & 0.9222 & 10.50\\
        RDBN + RIRN & \textbf{31.83} & \textbf{0.9227}& \textbf{30.17} & \textbf{0.9120} & \textbf{32.71} & \textbf{0.9322} & 8.53\\
        \midrule
        GRU & 27.53 & 0.8335 & 25.11 & 0.7890 & 26.69 & 0.7956 & 22.24\\
        GRU + RIRN & \textbf{28.58} & \textbf{0.8773} & \textbf{26.36} & \textbf{0.8217} & \textbf{28.60} & \textbf{0.8428} & 20.94\\
        \midrule
        LSTM & 26.94 & 0.8365 & 25.22 & 0.7948 & 26.87 & 0.8046 & 19.14\\
        LSTM + RIRN& \textbf{29.16} & \textbf{0.8813} & \textbf{27.24} & \textbf{0.8400} & \textbf{29.12} & \textbf{0.8584} & 16.90\\
        \bottomrule
    \end{tabular}
\end{table}
In Figure~\ref{fig:gopro_comparison}, \ref{fig:reds_comparison} and \ref{fig:dvd_comparison}, we show the visual effect of our RIRN-applied results on various RNNs on GOPRO~\cite{Nah_2017_CVPR}, REDS~\cite{Nah_2019_CVPR_Workshops_REDS}, and DVD~\cite{Su_2017_CVPR} datasets.
By applying RIRN, the textures are better deblurred and better recognizable compared with the results obtained from the baseline methods.
{\revision{
We also show the generalizability of our RIRN on the real blurry videos~\cite{cho2012registration} in Figure~\ref{fig:real}.
}}




\section{Conclusion}
\label{sec:conclusion}

In this paper, we proposed a new method to improve the existing recurrent neural networks with our recurrence-in-recurrence network architecture.
The recurrence-in-recurrence network consists of inner-recurrence module and the adaptive temporal blending method to augment the baseline recurrent networks.
The inner-recurrence module learns to model the temporal variation of hidden states and provides a complementary information that is often unseen at the previous time step by looking into further past frames.
The adaptive temporal blending of features can selectively extract necessary information and suppress unwanted part of the information.
We show that our RIRN consistently improves the video deblurring performance on various RNN-based methods and exhibits state-of-the-art performance.



\section{Acknowledgment}
\label{sec:acknowledgment}
This work was supported by IITP grant funded by the Korea government(MSIT) [No.2021-0-01343, Artificial Intelligence Graduate School Program (Seoul National University)]

\clearpage
\bibliography{main}
\end{document}